# Three Key Questions on Fractal Conductance Fluctuations: Dynamics, Quantization and Coherence

A.P. Micolich [1,2,*], R.P. Taylor[2,*], T.P. Martin[2], R. Newbury[1], T.M. Fromhold[3], A.G. Davies[4], H. Linke[2], W.R. Tribe[4], L.D. Macks[1], C.G. Smith[4], E.H. Linfield[4] and D.A. Ritchie[4]

[1] *School of Physics, University of New South Wales, Sydney NSW 2052, Australia.*

[2] *Physics Department, University of Oregon, Eugene OR 97403-1274.*

[3] *School of Physics and Astronomy, University of Nottingham, Nottingham NG7 2RD, U.K.*

[4] *Cavendish Laboratory, University of Cambridge, Cambridge CB3 0HE, U.K.*

Recent investigations of fractal conductance fluctuations (FCF) in electron billiards reveal crucial discrepancies between experimental behavior and the semiclassical Landauer-Buttiker (SLB) theory that predicted their existence. In particular, the roles played by the billiard's geometry, potential profile and the resulting electron trajectory distribution are not well understood. We present new measurements on two custom-made devices – a 'disrupted' billiard device and a 'bilayer' billiard device – designed to directly probe these three characteristics. Our results demonstrate that intricate processes beyond those proposed in the SLB theory are required to explain FCF.

Conductance fluctuations have proven to be a sensitive probe of electron dynamics and chaotic phenomena in semiconductor billiards. These billiards consist of electrons scattering ballistically around a micron-sized two-dimensional (2D) cavity bounded by shaped walls.[1,2] Billiards are typically defined in the 2D electron gas (2DEG) of an AlGaAs/GaAs heterostructure using surface gates,[3] resulting in a 'soft' potential profile with approximately parabolic walls and a flat bottom (see Fig. 1).[4] At milli-Kelvin temperatures, quantum interference dominates the electrical conductance of the billiard, generating reproducible fluctuations as a function of magnetic field (see Fig. 2).[1,2] In 1996, a semiclassical Landauer-Buttiker (SLB) theory[5] was used to predict that soft-walled billiards support 'mixed' chaotic/stable electron dynamics, leading to fractal conductance fluctuations (FCF) that exhibit recurring structure at increasingly fine magnetic field scales,[6] and which have since been observed experimentally.[7,8] A number of theoretical studies have followed Ref. 5 proposing alternative and sometimes contradictory explanations for fractal conductance fluctuations (FCF). These new theories include a semiclassical analysis based on the Kubo formalism,[9] a quantum-mechanical analysis of both the fully chaotic[10] and integrable[11] regimes, and a 2D tight-binding model.[12] The focus of recent work is to inspire a more complete understanding of this phenomenon by exploring the roles of dynamics, quantization and coherence in generating FCF. This has been achieved both by using novel low-T STM techniques,[13] and in our case, by devising experiments that target the key features differentiating the existing theoretical models.

In this paper, we present three new experiments, designed to *directly* target key differences between the contending theories for FCF.[5,9-12] In particular, we address the fundamental question of the link between FCF and the underlying electron trajectory distribution. First, using a 'disrupted' billiard device (Fig. 1(a)), we explore the effect of altering the geometry whilst maintaining a constant confining potential profile and observe that the resulting change in trajectories has little effect on the statistical properties of the FCF. Second, we reduce the phase coherence time by increasing the temperature in order to systematically eliminate the contribution of longer trajectories in this device, and find that the FCF do not respond in the manner expected from the original SLB theory. Finally, using a 'bilayer' billiard device (Fig. 1(b)) we vary the potential profile whilst keeping geometry constant to test the predicted critical link between profile and electron dynamics,[5] and find that changes in potential profile only have a measurable effect on the statistics of FCF in the regime where semiclassical theories are valid.

A key statistical parameter in the study of FCF is the fractal dimension $D_F$, which quantifies the scaling relationship between conductance fluctuations at different scales.[5,7-13] An important prediction of the semiclassical Kubo theory[9] compared to the SLB theory[5] is that, although the FCF should be affected by the electron dynamics and softness of the potential profile, $D_F$ should be independent of the detailed geometric shape of the billiard. Two other recent theories go further and suggest that the existence and properties of FCF may not depend on *any* of these three parameters. The first reports fractal fluctuations in a strongly quantized non-chaotic billiard[11] where a soft-wall profile and its associated mixed phase-space do not occur. The second reports fractal fluctuations in 2D tight-binding models of both chaotic and non-chaotic billiards[12], where FCF occur *without* mixed electron dynamics and the potential profile plays no role.

We recently reported an experimental study[14] of the $D_F$ dependence on tunable parameters such as the enclosed billiard area $A$, temperature $T$ and the number of conducting modes $n$ in the entrance and exit quantum point contacts (QPCs). Remarkably, we found that the $D_F$ of the FCF is directly dependant on an empirical parameter $Q$ that quantifies the resolution of the billiard's energy level spectrum. This parameter $Q$ is defined as the ratio of the billiard's mean energy level spacing $\Delta E_S$ to the billiard's mean energy level broadening $\Delta E_B$. The broadening $\Delta E_B$ is affected by the quantum lifetime $\tau_q$, which is limited by phase breaking scattering. The relationship between $D_F$ and $Q$ was discovered through the observation that all of the measured $D_F$ values condensed onto a single, well-defined curve as a function of $Q$ (see Fig. 3(a) – solid symbols), referred to as the '$Q$ curve'. The evolution of FCF charted on the $Q$ curve spans a large range of billiard parameters. The fluctuations are fractal over the entire range between the limits $Q \to 0$ and $Q \cong 10$ where $D_F \to 1$ and the fluctuations become non-fractal. Starting at $Q = 0$ in Fig. 3(a), $D_F$ rises sharply with increasing $Q$, attaining a peak value of ~1.5 at exactly $Q = 1$, and thereafter decreases linearly with increasing $Q > 1$. In this paper, we use the '$Q$ curve' discovered in Ref. 14 as a tool to answer several unresolved questions related to the physics of FCF.

The disrupted billiard device overcomes a key limitation in Ref. 14, which is that the similar geometries of the billiards used – all were 'empty' (no scattering obstacles in the billiard), rectangular billiards – leads to electron trajectory distributions sufficiently similar as to escape detection as an influence on $D_F$. Figure 1(a) shows a scanning electron micrograph of the gate pattern for the disrupted billiard device, which consists of two billiards, each formed by three independently controllable surface-gates. The empty square billiard on the left (gates 1, 2 and 3) is 1 μm wide, with two QPCs (bottom left corner)[15] and serves as a control device for the experiment. The billiard on the right (gates 4, 5 and 6) is the disrupted billiard – a square nominally identical to that on the left, but with the addition of a narrow, diagonal, trajectory disrupting 'finger gate' that extends from the corner between the QPCs to the billiard's center. This finger gate is designed to radically alter the electron trajectories with minimal impact on both the overall geometry established by the outer walls (a square billiard) and the enclosed billiard area $A$ (the finger-gate reduces $A$ by < 3%). The devices are located within close proximity (< 1 μm apart) on the same chip to ensure closely matched material parameters (electron density $n_s$ = 4.2 × $10^{11}$ $cm^{-2}$ and mobility μ = 2.5 × $10^6$ $cm^2$/Vs) and measurement conditions (e.g., equal $T$). For each billiard, we tune the gate biases so that their two QPCs both transmit either $n$ = 2 or 5 modes each. The combination of proximity and identical $T$ and $n$ ensures that between the two devices the measured $\tau_q$ differs by < 10% and hence $Q$ differs by < 2.5% for each data set at a given $T$ and $n$. Devices were mounted in thermal contact with the mixing chamber of a dilution refrigerator and measured using a low frequency, constant current lock-in technique.[16] Fractal analysis of the conductance fluctuations was performed using a modified box-counting method.[16,17]

The two traces in Fig. 2(a) show the measured conductance for the empty and disrupted billiards at $T$ = 50 mK and $n$ = 5, and reveal FCF superimposed on a smoothly varying classical background. We isolate this background using a locally weighted least squares fitting procedure. These fits are shown as dashed lines in Fig. 2(a) and are qualitatively similar to traces measured at $T$ = 4.2 K, where quantum interference fluctuations are heavily suppressed, supporting the validity of the fits. The two background fits are significantly different, demonstrating that the finger gate has altered the electron trajectory distribution in the billiard. This is further confirmed by the fact that the empty billiard has a larger overall conductance than the disrupted billiard; the finger gate acts to obstruct direct trajectory paths between the two QPCs.[18] In order to facilitate a direct comparison of the individual features of the two sets of fluctuations, we have subtracted the fitted backgrounds from each trace and overlaid them in Fig. 2(c). An inspection of these overlaid traces reveals the expected clear differences in the individual fluctuation features. However, despite these differences, the fractal statistics for the two traces, as quantified by $D_F$, are effectively identical. This is demonstrated in the $Q$ curve in Fig. 3(a), where the data for both billiards condense onto the original curve found in Ref. 14. This result clearly demonstrates that, in terms of determining $D_F$, only $Q$ is important. *In particular, provided the enclosed area is the same for the two billiards, its geometry and the resulting detailed nature of the electron trajectories do not determine $D_F$.* This insensitivity of $D_F$ to geometric details contradicts the SLB theory[5] and agrees with Ref. 9.

We now highlight further discrepancies between experimental behavior and the SLB theory.[5] The region where semiclassical theories[5,9] are valid happens to center on $Q = 1$. In this regime $\tau_q$ is sufficiently long that typical electron waves traverse the billiard without suffering phase-breaking scattering, and the ratio $S$ of the billiard width to the electron Fermi wavelength is sufficiently large (~25) for the semiclassical picture of wave propagation along classical trajectories to hold. Significantly, $Q = 1$ coincides with the peak in the $Q$ curve and the peak $D_F$ of ~1.5 matches the maximum value predicted by the SLB theory.[5,8] We now examine how the FCF evolves as $Q$ moves away from unity. Consider, first, reducing $Q$ below 1, which we achieve through a reduction of $\tau_q$. According to Ref. 5, $D_F$ is directly related to the exponent $\gamma$ of a power-law distribution of the areas enclosed by closed trajectory loops. Therefore, $D_F$ should depend only on parameters that directly affect $\gamma$ through rearrangements of the area distribution, and hence should not depend on parameters that determine $\tau_q$ such as $T$. Instead, reducing $\tau_q$ should simply render the longer trajectories phase-incoherent and prevent the largest enclosed areas from contributing to the FCF. Thus, fluctuations with small magnetic field period $\Delta B$ should be suppressed first, leaving the large $\Delta B$ fluctuations relatively unaffected.[1,5] In Fig. 4, we show scaling plots obtained from the disrupted billiard (Fig.1 – right) for $T = 50$ mK (top), 500 mK and 1.2 K (bottom). Not only does the whole $\Delta B$ spectrum evolve with $T$, maintaining the fractal scaling relationship and leading to a change in $D_F$ that depends on both $\tau_q$ and $T$, but *the* *lower ΔB cut-off in fractal scaling shifts in the opposite direction to that predicted by the simple SLB theory arguments above.* We also observe this effect in the empty square billiard (Fig. 1 – left). Interestingly, the semiclassical Kubo theory[9] agrees with the experimentally observed $D_F$ evolution with $T$ and $\tau_q$. We observe a similar behavior for $Q > 1$. This is achieved by increasing $\Delta E_S$ through a reduction of billiard area. For smaller billiard areas, the Heisenberg time $t_H = \hbar/\Delta E_S$ is reduced, preventing the longer trajectory loops from contributing to the semiclassical process, and so suppressing the small $\Delta B$ fluctuations. In contrast to this SLB prediction[5], we find that the *whole* $\Delta B$ spectrum evolves to maintain the fractal scaling relationship, similar to the $Q < 1$ case. In summary, in moving away from $Q = 1$ (whether by increasing or decreasing $Q$) we find that the $\Delta B$ range over which the FCF are observed does not decrease, contradictory to the behavior predicted by the SLB theory.[5] Instead, the fractal character is preserved and $D_F$ evolves smoothly, decreasing gradually towards 1. A consequence is that the FCF are observed over substantially larger ranges of $Q$ than predicted, persisting well beyond the range of conditions required for the semiclassical theories to be valid. This behavior is, however, consistent with aspects of the other theoretical studies,[10-12] which indicate that FCF can exist for high $Q$,[11] well outside the conditions required by Refs. 5,9. However, at present, a detailed explanation for our observation of FCF at *both* high and low $Q$ is lacking.

We now turn to the role of potential profile in determining $D_F$. This is achieved using the bilayer billiard device shown schematically in Fig. 1(b), which features a pair of parallel, closely spaced 2DEGs at depths of 90 nm (shallow) and 140 nm (deep) beneath the heterostructure surface. The concept, architecture, fabrication and initial characterization of this device are detailed in Ref. 18, and here we use this device to study the relationship between FCF and potential profile. In brief, a single set of three surface gates with a geometry identical to the left-hand (empty) billiard in Fig. 1(a) are used to define both the deep and shallow billiards, which, as a result, have the same nominal geometry but differing potential profile by virtue of their different depths beneath the surface gates. Using a selective gating technique,[16,18] the billiards can be measured independently in a two-step process. The 2DEGs have matched $n_s = 2.9 \times 10^{11}$ cm$^{-2}$, similar mobilities ($\mu = 1.3 \times 10^6$ (shallow) and $1.1 \times 10^6$ cm$^2$/Vs (deep)) and identical $T$. For both billiards, the gates are tuned so that both QPCs have matching $n$ = 2, 5 or 8. Under these conditions, the two billiard areas $A$ differ by < 15%.[18] Based on the data in Ref. 14, we predict that this difference in $A$ produces less than a 1% change in $D_F$. In terms of geometry ($n$, $A$ and gate shape), the two billiards are essentially identical.[18] To determine the profiles of the two billiards, we used a self-consistent Schrödinger-Poisson model.[4] The shallow billiard has the softer profile due to the smaller gate bias required to define it.[18] The potential gradient at the Fermi energy (used as a measure of softness) differs by a factor of three between the two billiards and the two profiles differ by ≥ 0.5 meV (corresponding to 5% of $E_F$) across more than a quarter of the width of the billiard. Given the predicted critical sensitivity to profile,[5] this difference is expected to significantly impact on the details of the FCF predicted by the SLB and Kubo theories.[5,9]

Typical FCF for the shallow and deep billiards are shown in Fig. 2(b). A procedure identical to that employed for the traces in Fig. 2(a) and 2(c) is used to produce the fitted backgrounds in Fig. 2(b) and the overlay of the background-subtracted FCF in Fig. 2(d). The two classical background fits in Fig. 2(d) are strikingly similar, confirming that the shallow and deep billiards have the same nominal geometry (i.e., size and shape). Figure 2(d) demonstrates that the differing wall profile induces a significant change in the precise details of the FCF as expected if the dynamics have strong dependence on profile, and intuitively, one might expect that the fractal scaling has changed, as predicted by the semiclassical theories.[5,8,9] In Fig. 3(b), the $D_F$ vs. $Q$ data obtained for the deep and shallow billiards is superimposed on the original data from Ref. 14. Two separate results are indicated in this new data. In the regime $Q < 1$, the $D_F$ vs. $Q$ behavior of the deep and shallow billiards agree well within experimental uncertainties. In other words, although the difference in billiard profile is sufficient to induce changes in the individual features of the FCF, the statistical characteristics of the FCF are not affected. *In the vicinity of $Q = 1$, however, the $D_F$ values obtained for the deep billiard (with the harder wall profile) are significantly lower than those measured for the shallow billiards.* Here, the conditions required by the semiclassical theories[5,8,9] are satisfied, and $D_F$ is observed to be sensitive to changes in potential profile, in good agreement with theory.[5,9] The dashed line in Fig. 3(b) indicates a predicted trend for the deep billiard, where the change to a harder potential profile suppresses the peak $D_F$ whilst maintaining the general form of the $Q$ curve. Unfortunately, high $Q$ measurements for the deep billiard were not possible due to difficulties in defining small billiards with controlled QPCs on the bilayer heterostructure.

In conclusion, we have presented targeted experiments on two new devices aimed at directly probing the impact of billiard geometry and soft-wall profile on the properties of FCF to better understand the role of dynamics, quantization and coherence in generating FCF. We found that $D_F$ is unaffected by the change in geometry induced by the introduction of a trajectory disrupting 'finger gate'. This insensitivity contradicts the SLB theory for FCF[5] but is in general agreement with other recent theories.[9-12] The role of potential profile depends on whether semiclassical conditions exist within the billiard. In the vicinity of $Q = 1$, $D_F$ is sensitive to the potential profile, in agreement with Refs. 5 and 9. However, for $Q < 1$, where the semiclassical approximation breaks down, $D_F$ is insensitive to profile; no theory for FCF currently exists for this regime. Our results suggest that more complicated processes than those predicted in the semiclassical models are responsible for the observed behavior of FCF.

We thank M. Pepper, P.E. Lindelof and A.R. Hamilton for support and valuable discussions. APM is an ARC Postdoctoral Fellow, RPT is a Research Corporation Cottrell Scholar and TPM is an NSF IGERT Fellow. We acknowledge financial support from the Australian Research Council.

[18] FFTs for the two sets of FCF data show significant differences particularly for high $\Delta B$, demonstrating the finger gate is active.

[19] A.P. Micolich, R.P. Taylor, A.G. Davies, T.M. Fromhold, H. Linke, L.D. Macks, R. Newbury, A. Ehlert, W.R. Tribe, E.H. Linfield and D.A. Ritchie, Appl. Phys. Lett. **80**, 4381 (2002).

Figure Captions

Figure 1: (a) Scanning electron micrograph of the disrupted billiard device showing the surface gates (numbers discussed in text) that define the empty (left) and disrupted (right) billiards. (b) Schematic (not to scale) of the bilayer billiard device where a common set of gates defines billiards in shallow and deep 2DEGs. The relative potential profiles of these two billiards are discussed in the text and are purely illustrative.

Figure 2: FCF (bottom axis) for: (a) the empty (upper) and disrupted (lower) billiards and (b) the shallow (upper) and deep (lower) billiards. The dashed lines are fits to the classical background. An overlay of FCF traces with classical background fits subtracted and $B > 0$ (top axis) for: (c) the empty (thin line) and disrupted (thick line) billiards with $n = 5$ and $T = 50$ mK and (d) the shallow (thin line) and deep (thick line) billiards with $n = 2$ and $T = 50$ mK.

Figure 3: (a) The $D_F$ values from the empty and disrupted billiards as a function of $Q$ overlaid on the original $Q$ curve from Ref. 14. (b) The same data as (a) with the addition of the shallow and deep billiard $D_F$ vs. $Q$ values. The dashed lines are guides to the eye, and error bars indicate the expected maximum uncertainty in $D_F$ and $Q$.

Figure 4: Fractal scaling plots for FCF data obtained from the disrupted billiard for $T = 50$ mK (top), 500 mK and 1.2 K (bottom). The linear fits yield $D_F$ values of 1.51, 1.43 and 1.20 respectively. Arrows indicate the lower cutoffs for fractal behavior, which occur at log $\Delta B = -2.09$, $-2.18$ and $-2.43$ respectively. The expected uncertainty on these lower cut-offs is indicated by the error bar.

A.P. Micolich *et al.*
"Three Key Questions on Fractal Conductance…"
Physical Review B
Figure 1

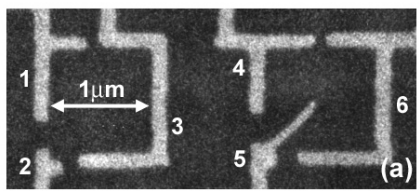

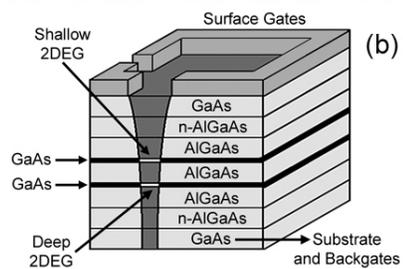

A.P. Micolich *et al.*
"Three Key Questions on Fractal Conductance…"
Physical Review B
Figure 2

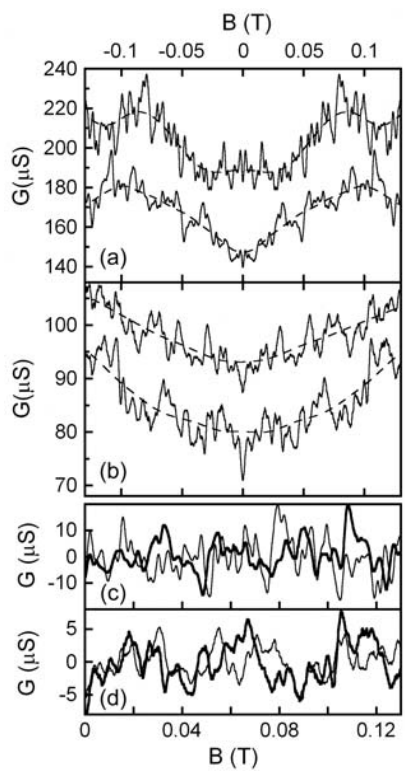

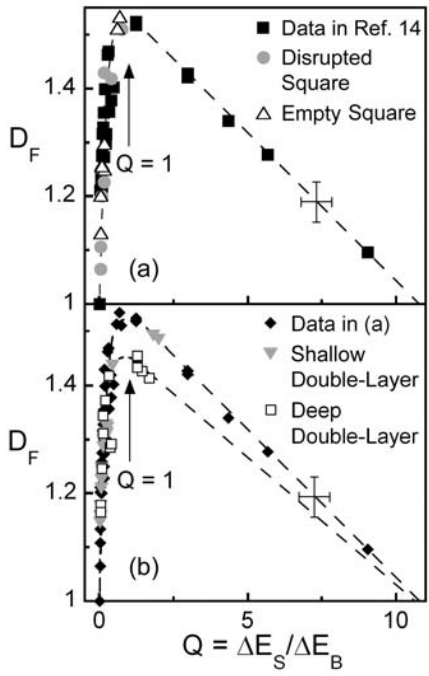

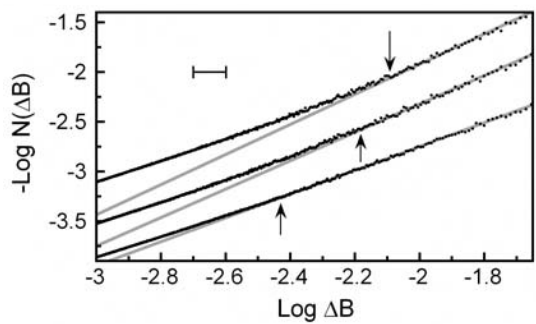